\documentstyle[psfig]{mn}

\def\spose#1{\hbox to 0pt{#1\hss}}
\def\simlt{\mathrel{\spose{\lower 3pt\hbox{$\mathchar"218$}}
     \raise 2.0pt\hbox{$\mathchar"13C$}}}
\def\simgt{\mathrel{\spose{\lower 3pt\hbox{$\mathchar"218$}}
     \raise 2.0pt\hbox{$\mathchar"13E$}}}

\def\etal{et~al. }
\def\eg{e.g. }

\def\hmpc{\;h^{-1}{\rm Mpc}}

\def\kms{{\rm km\;s}^{-1}}
\def\kmsmpc{\kms\;{\rm Mpc}^{-1}}

\def\oii{O[{\sc II}]\,}

\def\refer{\par \noindent \hangindent=3pc \hangafter=1}

\def\AJ{A.J.}
\def\ApJ{Ap.J.}

\def\MN{MNRAS}

\def\Nature{Nature}

\begin{document}

\title[The Clustering of Hot and Cold IRAS Galaxies: The Redshift
Space Correlation Function]{The Clustering of Hot and Cold IRAS
Galaxies: The Redshift Space Correlation Function}

\author[Hawkins et al.] 
{E.Hawkins$^{1}$\footnotemark, S.Maddox$^{1}$, E.Branchini$^{2}$ and
W.Saunders$^{3,4}$\\ 
$^{1}${School of Physics and Astronomy, University of Nottingham,
Nottingham, NG7 2RD, UK.} \\  
$^{2}${Dipartimento di Fisica, Universita' degli Studi Roma Tre,     
Via della Vasca Navale 84, I-00146 Roma, Italy} \\
$^{3}${Royal Observatory, Blackford Hill, Edinburgh, EH9 3HJ, UK.}\\
$^{4}${Anglo-Australian Observatory, P.O. Box 296 Epping NSW 1710
Australia }\\
}

\maketitle

\date{}

\begin{abstract}
We measure the autocorrelation function, $\xi$, of galaxies in the
{\it IRAS} Point Source Catalogue galaxy redshift (PSCz) survey and
investigate its dependence on the far-infrared colour and absolute
luminosity of the galaxies. We find that the PSCz survey correlation
function can be modelled out to a scale of $10\hmpc$ as a power law of
slope $1.30 \pm 0.04$ and correlation length $4.77 \pm 0.20$. At a
scale of $75 \hmpc$ we find the value of $J_3$ to be $1500 \pm 400$. 

We also find that galaxies with higher $100\mu m/60\mu m $ flux ratio,
corresponding to cooler dust temperatures, are more strongly clustered
than warmer galaxies. Splitting the survey into three colour
subsamples, we find that, between 1 and $10 \hmpc$, the ratio of $\xi$
is factor of 1.5 higher for the cooler galaxies compared to the hotter
galaxies.  This is consistent with the suggestion that hotter galaxies
have higher star-formation rates, and correspond to later-type
galaxies which are less clustered than earlier types.

Using volume limited sub-samples, we find a weak variation of $\xi$ as
a function of absolute luminosity, in the sense that more luminous
galaxies are less clustered than fainter galaxies. The trend is
consistent with the colour dependence of $\xi$ and the observed
colour-luminosity correlation, but the large uncertainties mean that
it has a low statistical significance.

\end{abstract}


\section{Introduction}
\label{intro}
\footnotetext{e-mail:ppxeh@nottingham.ac.uk}

The relation between the distribution of galaxies and the distribution
of mass is now one of the most important problems in large-scale
structure. Empirically, it has been found  that different galaxy
types have different clustering amplitudes and hence that galaxies are
biased tracers of the mass distribution. Evidently, we need to
understand the physical mechanisms responsible for these biases if we
are to establish the connection between galaxy tracers and the mass.
This problem can be understood in terms of two related questions:
(i) How does the clustering of a galaxy sample depend on the 
properties of the galaxies? and (ii) How do the properties of galaxies
depend on their local environment?
 
Many ideas have been put forward as to how the environment may modify
galaxy properties, from schematic ideas concerning feedback in the
formation process (White \& Rees 1978; Dekel \& Rees 1987) to
hydrodynamical simulations of galaxy formation incorporating cooling
and dissipation (e.g. Katz \& Gunn 1991; Navarro \& White 1993; Pearce
\etal 1999).
Over the next few years we can expect major advances in hydrodynamic
simulations employing parallel computers and there is a real prospect
of understanding the relationship between galaxy morphologies and
environment within the context of specific theories of cosmological
fluctuations. We can hope that such numerical simulations will eventually
be able to provide detailed predictions of the spatial distribution of 
galaxies as a function of their stellar content, rotation velocity,
star-formation rate, and morphological type.

The simplest bias schemes are those based on the statistics of high
density peaks (Kaiser 1984, Mo \& White 1996). 
Since the optical magnitude of a galaxy is correlated
to the depth of the potential via the Tully Fisher relation for
spirals and $D_n-\sigma$ relation for ellipticals, this would suggest 
that the brightest galaxies will be more strongly correlated
than fainter galaxies. 
Realistic galaxy formation models rely on feedback mechanisms to limit
star-formation, either acting internally within each galaxy or
involving interactions with other galaxies and the intergalactic medium. So
galaxies in regions of high local galaxy density are likely to have a
reduced star-formation rate (SFR).  Indeed this is confirmed by the
observation that later-type optical galaxies, which have higher SFRs,
have a much lower clustering amplitude than earlier galaxy types (\eg
Rosenberg, Salzer \& Moody 1994; Loveday \etal 1995; Loveday, Tresse
\& Maddox 1999).

Galaxies with a high SFR tend to be bright in the far infra-red (FIR)
because of the thermal emission from dust heated by young stars.  Thus
the observations that {\it IRAS} galaxies (selected on their $60 \mu$m
flux) tend to avoid rich galaxy clusters, and have a lower clustering
amplitude are consistent with this general picture.  In this paper we
divide the PSCz sample of {\it IRAS} galaxies into subsamples based on
their FIR colour. Since the galaxies with warmer FIR colours
are those with higher SFR, we can further test the dependence on
star-formation rate. Mann, Saunders \& Taylor (1996) selected warm and
cool sub-samples of {\it IRAS} galaxies from the QDOT survey based on
their 60$\mu m$/100$\mu m$ flux ratio, and found marginal evidence
that warmer galaxies are more strongly clustered than cooler galaxies,
the opposite to what is expected from the morphology density
relation.  However they did not consider the result very significant
compared to the expected cosmic variance.

Using optical galaxy samples, a number of authors have found
indications of changes of the clustering amplitude with galaxy
luminosity, (\eg Valls-Gabaud, Alimi \& Blanchard 1989; Park \etal
1994; Moore \etal 1994; Loveday \etal 1995; Benoist \etal 1996) but
the samples are small and the observed amplitude shifts do not have a
high statistical significance.  The expected luminosity dependence for
{\it IRAS} galaxy samples is less clear, since the FIR luminosity of a
galaxy is likely to depend on both the mass of its dark halo and its
specific SFR.  If the FIR luminosity is correlated to the halo mass,
then brighter galaxies should have a higher clustering amplitude. On
the other hand, a high SFR will make a galaxy more luminous, and
galaxies with a high star-formation rate tend to have weaker
clustering.  It is not easy to predict which of these is the dominant
effect.  Observationally Szapudi \etal (2000) and Beisbart \& Kerscher
(2000) have used mark correlation functions with volume limited
sub-samples to examine the luminosity dependence of clustering in the
PSCz survey. Over the narrow range of luminosities they consider they
concluded that there is no significant luminosity dependence.

In this paper we investigate clustering amplitude for different
sub-samples of galaxies from the PSCz survey.  We begin by summarizing
the relevant aspects of the PSCz survey and describe various
sub-samples that we will study here. In Section~\ref{meth} we show our
methods and in Section~\ref{res} we give our results for the whole
sample and consider how the amplitude of the redshift space
autocorrelation function depends on the colour and luminosity
selection of the galaxies.  Finally, we discuss our conclusions in
Section~\ref{discuss}.

\section{The PSC Redshift survey} 
\label{PSCZ} 

\subsection{Description}

Galaxies for the PSCz were selected from the {\it IRAS} Point Source
Catalogue (Beichman \etal 1988) by positional, identificational flux
and colour criteria designed to accept almost all galaxies while
keeping contamination by Galactic sources to an acceptable level. The
significant differences compared with the QDOT survey (Lawrence \etal
1999), were that the colour 
criteria were relaxed to avoid excluding previously uncatalogued
galaxies with unusual (especially cool) colours and the sky coverage
is increased to 84 per cent by including areas of high cirrus
contamination. These changes increase the contamination of the galaxy
catalogue from local galactic sources (stars, planetary nebulae,
cirrus etc) but ensure the sample has a higher completeness. The
contaminating Galactic sources were excluded by a combination of {\it
IRAS} and optical properties (from sky survey plates), and where
necessary spectroscopy. Redshifts were taken from all available
published or unpublished sources; principally Huchra's ZCAT, the LEDA
database, the 1.2Jy and QDOT surveys. Also 4500 new redshifts were taken on
the Isaac Newton Telescope, the Anglo-Australian Telescope, the
Cerro-Tololo 1.5-m and Cananea 2.1-m telescopes.  In total the catalogue
contains 14677 redshifts and the median redshift is $\sim 0.028$.  The
uniformity and completeness of the catalogue are discussed by Saunders
\etal (2000).
 
\begin{figure}
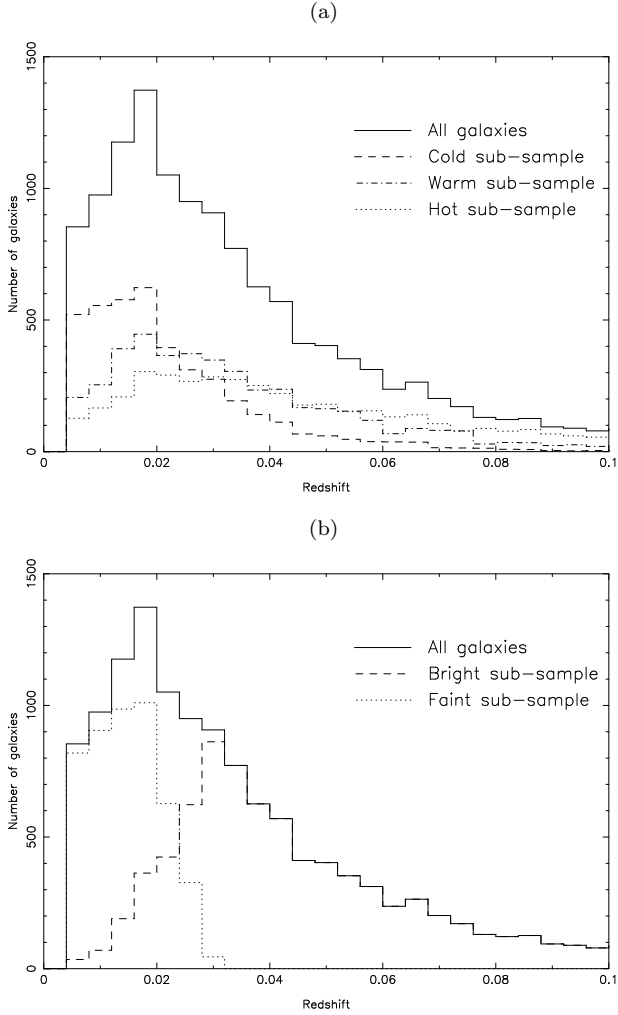

\begin{center} (a) \end{center}
\psfig{figure=./zfncol.ps,angle=-90,height=6cm,clip=}
\begin{center} (b) \end{center}
\psfig{figure=./zfnlum.ps,angle=-90,height=6cm,clip=}
\caption[]{ The redshift distributions for galaxy samples. (a) colour
selected sub-samples listed in Table~\ref{t_fits_col}. (b) absolute
luminosity selected sub-samples listed in Table~\ref{t_fits_lum}.
}\label{f_nz}
\end{figure}

The redshift distribution of the galaxies is shown in
Figure~\ref{f_nz}. It can be seen that the peak of the distribution is
near $z = 0.02$, and there is a tail which extends out to beyond $z = 0.1$.
This tail means that there are over 1200
galaxies in the survey with a redshift $>$ 0.08. 

For our present analysis, only galaxies with a redshift $>$ 0.004,
corresponding to a recession velocity of 1200 $\kms$ were selected, so
that local effects and the velocity uncertainties of around 120
$\kms$ were less significant.

\subsection{Sub-samples}
\subsubsection{Colour sub-samples} 
\label{col_samples} 

In order to investigate the dependence of clustering on the temperature
of the galaxies, we used the addscan $60\mu m$ and $100\mu m$ fluxes
($f_{60}$ and $f_{100}$) to sub-divide the parent redshift catalogue
into three sub-samples: hot galaxies, with $f_{100}/f_{60} < 1.7 $ ;
warm galaxies, with $1.7 < f_{100}/f_{60}< 2.3$; and cold galaxies,
with $f_{100}/f_{60} > 2.3 $. These boundaries correspond to
blackbody temperatures of around 31 and 28K, and were chosen to  
give roughly equal numbers in each sub-sample.  The actual numbers of
galaxies in the sub-samples were 4452, 4388 and 4107 respectively.
The mean $f_{100}/f_{60}$ colour ratios for the hot, warm and cold 
sub-samples were 1.31, 1.98 and 2.91 respectively, corresponding to
black-body temperatures of about 34K, 30K and 26.5K respectively. The
mean for the whole catalogue is 2.05, corresponding to a black-body
temperature of around 29.5K. 

\begin{figure}
\psfig{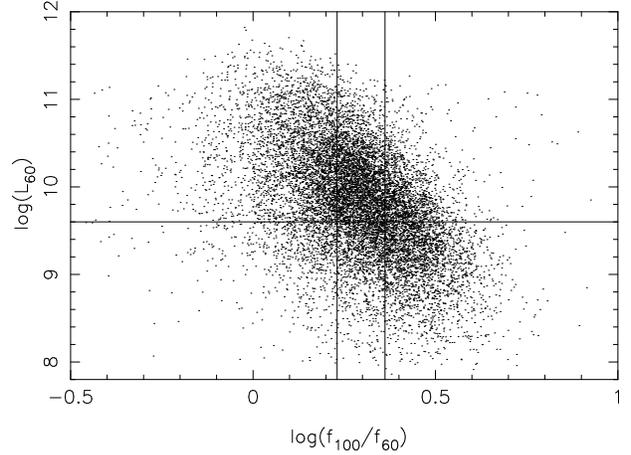}
\caption[]{Log plot of absolute luminosity at 60$\mu m$ against the $100
\mu m/60 \mu m$ colour. The vertical lines show the boundaries used
to select our colour sub-samples, and the horizontal line shows the
boundary between the luminosity samples. }
\label{f_colf}
\end{figure}

The redshift distributions for the sub-samples are plotted in
Figure~\ref{f_nz}(a) and it can be seen that the cooler samples tend
to peak at a slightly lower redshift than the hotter samples.  This
reflects the correlation between colour and absolute luminosity as
plotted in Figure~\ref{f_colf}: cooler galaxies tend to be fainter,
and so are only seen nearby.  Nevertheless, these differences in
redshift distributions are smaller than the difference for luminosity
selected sub-samples, as discussed in the next section.

\begin{figure*}
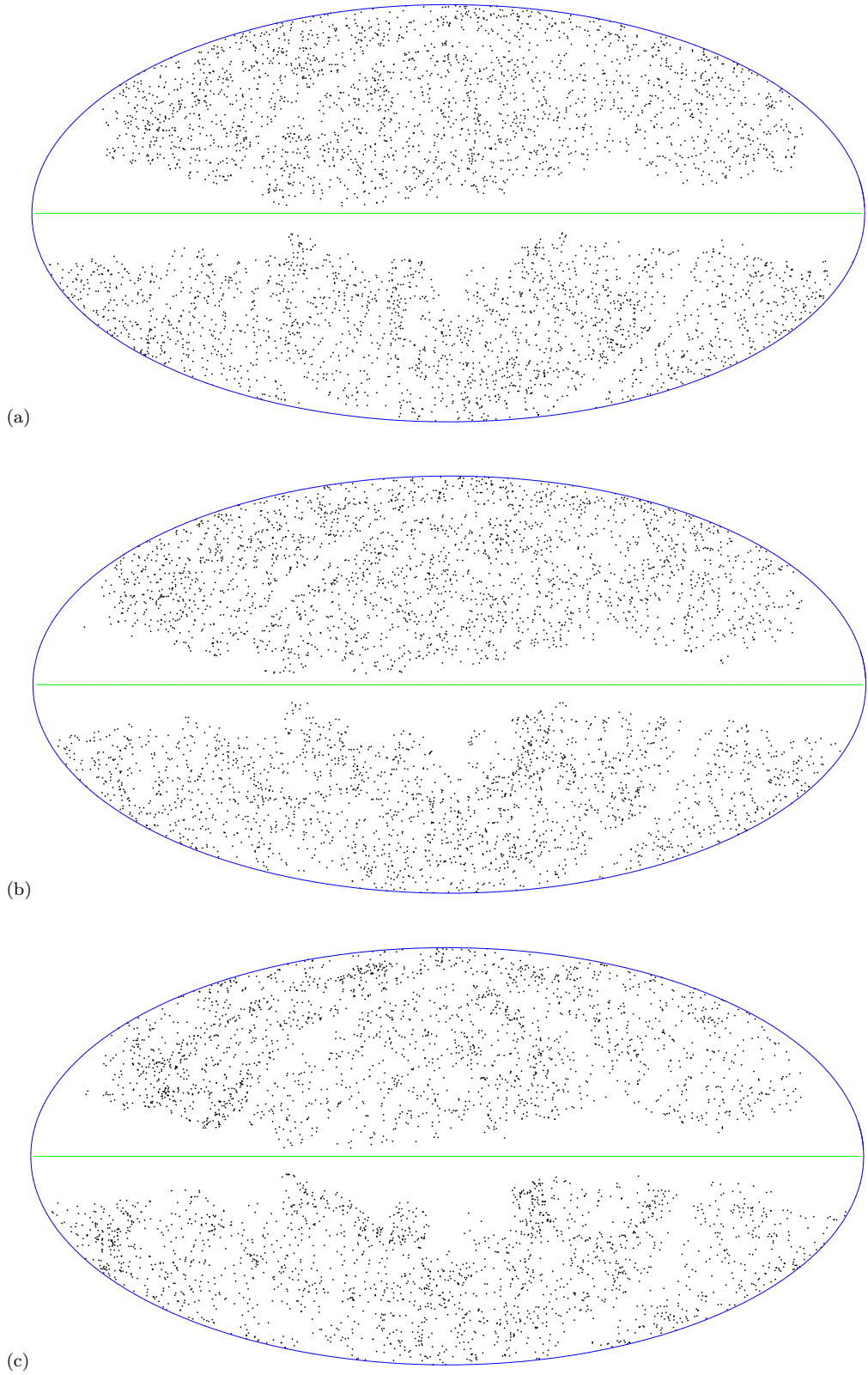

\begin{center}
\parbox{15.3cm}{
\hbox{(a){\psfig{figure=./skyplot_lt1.7.ps,angle=-90,height=6.5cm,clip=}}\vspace{0.7cm}}
\hbox{(b){\psfig{figure=./skyplot_1.7_2.3.ps,angle=-90,height=6.5cm,clip=}}\vspace{0.7cm}}
\hbox{(c){\psfig{figure=./skyplot_gt2.3.ps,angle=-90,height=6.5cm,clip=}}\vspace{0.7cm}}
}\caption[]{Plots of the position of galaxies on the sky for the three
colour sub-samples listed in Table~\ref{t_fits_col}: (a) hot, (b)
warm, and (c) cold. The line shows the Galactic equator and the
obscuration around this is masked out in the analysis. 
}\label{f_skyplots}
\end{center}
\end{figure*}

\begin{figure*}
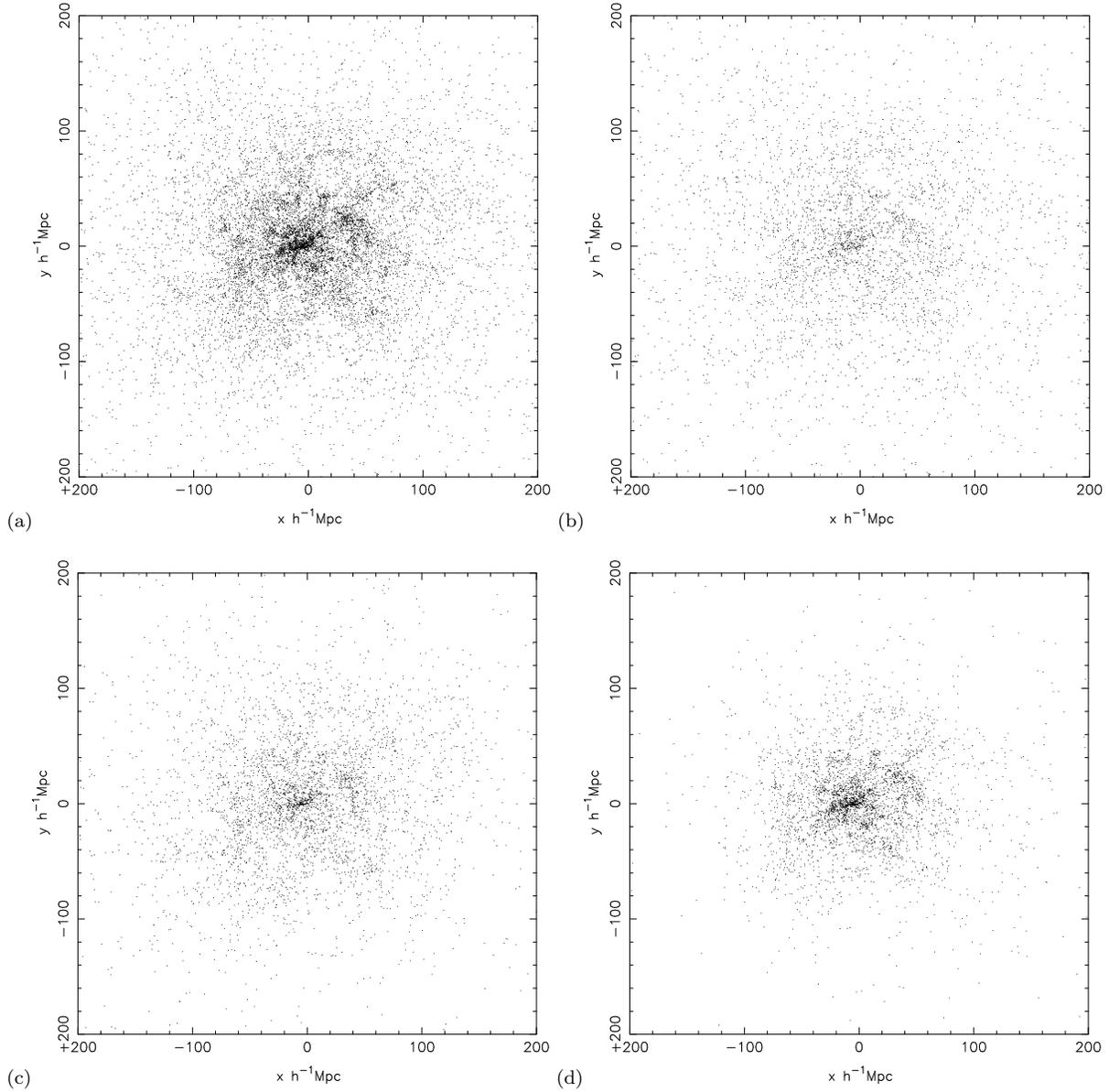

\begin{center}
\parbox{15.3cm}{
\hbox{(a){\psfig{figure=./fig_xy_all2.ps,angle=-90,height=8cm,clip=}}
      (b){\psfig{figure=./fig_xy_hot.ps,angle=-90,height=8cm,clip=}}}
\hbox{(c){\psfig{figure=./fig_xy_warm.ps,angle=-90,height=8cm,clip=}}
      (d){\psfig{figure=./fig_xy_cold.ps,angle=-90,height=8cm,clip=}}}
}
\caption[]{The galaxy positions projected onto the celestial equator
for the full survey and the three colour sub-samples listed in
Table~\ref{t_fits_col}: (a) all, (b) hot, (c) warm, and (d) cold. 
} 
\label{f_xyplots}
\end{center}
\end{figure*}

The angular distribution of these sub-samples are plotted in
Figure~\ref{f_skyplots}.  It can be seen that these samples show no
obvious gradients as a function of position on the sky, and also that
the cooler sample appears to be more strongly clustered.
Figure~\ref{f_xyplots} shows the projection of the galaxies on to a
plane along the celestial equator.  Again the cooler sample appears
more clustered than the warmer samples.  Though apparently quite
significant, these visual impressions should be treated with
caution. The cooler sample is shallower than the warmer samples, so
the angular clustering would be stronger, even if the samples had the
same spatial clustering. Also the cool sample has a higher space
density than the warmer samples, and this may enhance the visual appearance
of clustering. 

\subsubsection{Absolute luminosity sub-samples} 

The parent sample was also divided into sub-samples of absolute
luminosity at $60 \mu m$: faint galaxies, with $\log_{10}(L_{60}) <
9.6$; and bright galaxies with $\log_{10}(L_{60}) > 9.6 $, where
$L_{60} = 4 \pi d^2 f_{60}$ and $d$ is the luminosity distance to the
galaxy (we assumed $\Omega=1$ and $\Lambda=0$ to determine the
geometry, but at the distances considered here, cosmological effects
are negligible.) The units of $L_{60}$ are
$L_{\odot}h^{-2}.$\footnote{$h$ is the Hubble constant in units of
$100~\kmsmpc$ } Note that the usual definition of an ultra-luminous
{\it IRAS} galaxy corresponds to $\log_{10}(L_{60}) \simgt 11.4 $ (Soifer
et al 1987), and so the lower limit of our brighter sample is a factor
of $\sim$ 60 fainter than ultra-luminous galaxy samples.

The bright sample has a higher mean redshift, and hence covers a much
larger volume, but the galaxies are a more dilute sampling of the
density field and so a larger number is needed to give a similar
number of small separation pairs.  Hence the faint and bright samples
were chosen to have 4719 and 8228 galaxies respectively.

The redshift distribution for these sub-samples is shown in
Figure~\ref{f_nz}(b). Since the parent sample is flux limited, the low
luminosity galaxies have a sharply defined upper redshift limit
corresponding to the distance where the observed flux equals the flux
limit.  For the more luminous galaxies, the clustering signal is
dominated by the volume at higher redshifts. This means that the
clustering measurements for the two samples come from almost
independent volumes, and so cosmic variance introduces a large
uncertainty in the comparison.  The clustering measurements for colour
selected sub-samples are less affected by cosmic variance because they
cover very similar volumes, and so their comparison should be more
reliable.

\subsubsection{Volume limited sub-samples}

The cosmic variance between the luminosity sub-samples can be partly
overcome by using volume limited sub-samples. To do this, the parent
catalogue was split into sub-samples limited by distance and also by
absolute luminosity. Galaxies within a sphere of radius $x$, centred on
us, are included only if their absolute luminosity ($L_{60}$) is
greater than $4 \pi x^2 f_{min}$, where $f_{min}$ is the flux-limit of
the survey or sample (0.6 Jy for the full PSCz survey). Without galaxy
evolution these samples will be homogeneous.  Unfortunately the
numbers of galaxies in these samples, shown in Table \ref{t:vollim},
are rather small.  

\begin{figure}
\psfig{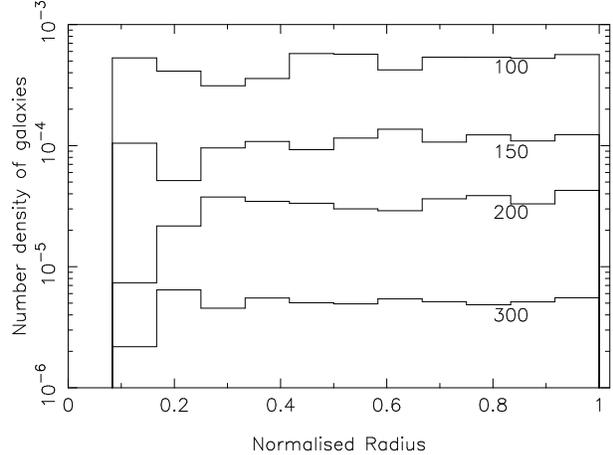} 
\caption[]{ Galaxy number density plotted against normalised radius
for the volume limited samples with radii labelled.  }\label{fv_nz}
\end{figure}

{\it IRAS} galaxies are known to show very rapid evolution (Saunders
et al 1990), and this could introduce a radial gradient in space
density in the volume-limited samples.  Figure~\ref{fv_nz} shows the
galaxy number density as a function of radius, normalised so that the
maximum radius of the volume is 1, for the samples with radius 100,
150, 200 and 300$\hmpc$.  No significant gradients are apparent, and
so we assume a constant density in our clustering estimates.  These
sub-samples allow another test for a luminosity trend in the
clustering as the larger volumes are limited to brighter galaxies than
the smaller volumes. Note that the different volumes have a
significant fraction of galaxies in common, so that the results for
different volume scales are not completely independent.

\subsection{PSCz mock catalogues}

Large N-body simulations from Cole \etal (1998) were used to create
mock PSCz surveys. The simulations  used the $AP^3M$ code of Couchman
(1991) loaded with $192^3$ particles of mass $1.64 \times
10^{12}\Omega_0h^{-1}M_{\odot}$ in a box with a co-moving size of
$345.6h^{-1}{\rm Mpc}$. More details can be found in Cole \etal
(1998). For this analysis three different cosmological models have
been considered. They are a spatially flat low density universe model
($\Lambda CDM$) with $\Omega _0 = 0.3$, $\Omega_{\Lambda} = 0.7$ and
$\Gamma = 0.25$, a spatially flat universe model ($SCDM\Gamma_{0.25}$)
with $\Omega _0 = 1.0$ and $\Gamma = 0.25$, and finally a spatially
flat universe model ($SCDM_{cobe}$) with $\Omega _0 = 1.0$ and $\Gamma
= 0.5$, normalised to match the amplitude of the COBE data. For pure
CDM-type models the shape parameter, $\Gamma = \Omega h$.

The galaxy catalogues were extracted from the numerical simulations by
first identifying a population of objects with similar properties to
the Local Group (LG). Then an observer is placed by considering the
observational constraints of peculiar velocity, local overdensity and
shear. All points within a sphere of $120h^{-1}{\rm Mpc}$ radius
around the observer are included, and the frame is rotated so that the
motion of the observer is the same as the LG peculiar velocity. Note
that this sphere radius is less than the size of the PSCz survey.  A
luminosity was randomly assigned to each galaxy in the mock
realizations to mimic the corresponding observational properties. The
PSCz density is reproduced by using the PSCz selection function to
reject points and also to randomly assign a flux to each point.
Unsurveyed regions are rejected using the PSCz mask to give the same
sky coverage as the real catalogue. Each mass point is associated with
a galaxy so the linear bias parameter, $b = 1.0$. Lastly the bias
against early-type galaxies in the original {\it IRAS} survey is introduced 
by rejecting an appropriate fraction of galaxies in clusters the cores
of clusters, which is meant to reproduce Dressler's morphology-density 
relation (Dressler 1980).
The resulting mock surveys have approximately the same correlation
function, redshift distribution and flux distribution as the real PSCz
sample.  

We calculated the correlation function for the 10
realizations of each model, and used the standard deviation about the
mean of $s_0$ and $\gamma$ to estimate of the uncertainties in these
measurements for the real sample.
These catalogues can also be split into the same bright and faint
sub-samples as the real catalogue.  Again, the standard deviation
between the realizations gives an estimate of the uncertainty on the
measurements for the real data, and so allows more reliable estimates
of the significance level of any variations seen in the real data.
Colours can be included in the mocks by considering
Figure~\ref{f_colf}, which we modelled as a linear relation between
$\log_{10}(f_{100}/f_{60})$ and $\log_{10}(L_{60})$ with a Gaussian
dispersion about the mean.  Each galaxy can then be assigned a colour
according to its luminosity, $\log_{10}(f_{100}/f_{60}) = A
\log_{10}(L_{60}) + B + C $, where $A=-0.117$, $B=1.427$ and $C$ is a random
number selected from a Gaussian distribution with variance of 0.2.  Then
colour sub-samples can be selected and tested in the same way as the
real data.
Since the luminosities were assigned at random to the mock galaxies,
and the colours are based on the luminosities, the clustering in the
mocks is independent of luminosity and colour.

\section{Methods for estimating the redshift space autocorrelation
function $\xi$}  
\label{meth}

There has been considerable interest in analysing the best way to
estimate the autocorrelation function from redshift surveys (\eg Landy
\& Szalay 1993; Vogeley \& Szalay 1996; Hamilton 1997a,b).
Reliable and unbiased estimates can be obtained by cross-correlating
the real data with an artificial unclustered catalogue which has the
same selection functions in angle on the sky and in redshift.  We
create this catalogue by generating random positions uniformly over
the sky, and rejecting positions behind the mask which defines the
boundaries of the real data.  For each position we then assign a
redshift from a distribution which matches the observed data.

We investigated several methods to generate the redshift distribution
for the random catalogues: random shuffling of the positions on the
sky relative to the redshifts; random shuffling followed by adding a
further random velocity from a Gaussian distribution with $\sigma =
500, 1000$ or $1500~\kms$; and generating a random redshift
distribution to match an analytic fit to the selection function
(Saunders \etal 2000).  For the full catalogue we found that random
shuffling with $1000~\kms$ smoothing and the analytic fit gave
indistinguishable results for $\xi$.  Rather than calculating a best
fit selection function for each sub-sample we simply generated
appropriate sub-samples using the $1000~\kms$ smoothed randomized
distribution. 
For each random catalogue we generated 12 times the number of random
points as real galaxies for the whole sample, and 20 times as many for
the sub-samples.

\subsection{Flux limited sample methods}

The estimate of $\xi(s)$ we used (Landy \& Szalay 1993) is given by
\begin{equation} \xi(s) = { \langle DD\rangle -2\langle DR\rangle + \langle
RR\rangle \over \langle RR\rangle},
\end{equation} 
where $\langle DD \rangle$ is the weighted sum over galaxy pairs with
separation $s$ 
in redshift space; $\langle RR \rangle$ is the weighted  sum of pairs
with the same separation 
in the random catalogue; and $\langle DR \rangle$ is the weighted sum
of galaxy-random pairs.  
In each of these sums, the pairs at separation $s$ are weighted by a factor 
\begin{equation} 
w_{ij} = \frac{1}{(1+4\pi n(z_i)J_3(s))(1+4\pi n(z_j)J_3(s))},
\end{equation} 
where $n(z_i)$ is the space density of observed galaxies at the
redshift $z_i$ for galaxy $i$, and $J_3 (s) = \int _0^s \xi(r)r^2
dr$ (Efstathiou 1988). 

To calculate $\xi(s)$ we first calculated $\xi(s)$ with a standard
pair weighting of the sample and fitted a power law to the results for
$1 < s < 10~h^{-1} {\rm Mpc}$. We then recalculated the correlation
function using $J_3$ from this best fit power law, truncated so that
the maximum $J_3$ is 1500.  It was found that only one iteration
was needed to produce a stable result with a consistent $J_3$ and
$\xi$. The calculation of $\xi(s)$ is relatively insensitive to the
precise form of the $J_3$ weighting employed. 
We then  fit the final correlation function with a power law of the form: 
\begin{equation}
\xi (s) = \left (\frac{s}{s_0} \right ) ^{-\gamma}.
\end{equation}

We calculate the correlation function for mock catalogues
in the same way as for the real sample.

\subsection{Volume limited sub-sample methods}
\label{sect_vollim} 
For the volume-limited sub-samples a different technique was used to
estimate the correlation functions. Following the method of Croft
\etal (1997), we used a maximum likelihood approach that assumes that
each distinct galaxy pair is an independent object.  This technique
uses the fact that $1+\xi(s)$ can be considered as the probability
distribution for galaxy pairs as a function of separation. Thus a
power law $\xi$ can be fitted directly to the galaxy pair distribution
without the need to specify arbitrary bins of pair separation. The
resulting power law parameters tend to be more robust than fitting to
a binned version of $\xi$, especially for small samples. 
The homogeneity of volume limited samples means that simple unit pair
weighting yields the best results.

If galaxy pairs were independent, the calculated likelihood contours
would directly provide 1-$\sigma$ uncertainties in the best-fit
parameters, equivalent to the Poisson error in the pair counts. Since
galaxy pairs are correlated, the effective number of pairs is reduced,
and so the uncertainties are increased by a factor $(1+4\pi n
J_3)$. These corrected Poisson estimates still do not allow for any
uncertainty introduced by the variance in mean density on the scale of
sample size. We used a simple approach to include this effect, by
considering the variance in the mean density within the sample volume,
${\rm var}(\delta n/n) = 3J_3 /R^3 $, where $R$ is the radius of the
volume. This introduces an uncertainty in $\xi$, which we add in
quadrature to the corrected Poisson estimates.

For volumes with radius $R \le 100\hmpc$ we have also analysed the ten
$\Lambda CDM$ mock catalogues using the same method. The $\Lambda CDM$
model was chosen because its clustering amplitude is the closest match
to the observed amplitude. The standard deviation between results
provides an independent estimate of the uncertainties in $s_0$ for
each sample. For the volumes we consider (with radius $R\ge 50 \hmpc$)
the value of $J_3$ is roughly constant, and we find that the two
methods agree to $\sim$10\% if we choose $J_3=1500$.  For volumes with
radius $R > 100\hmpc$ the mock catalogues are too small, and so we
have to rely on the analytical approximation to estimate the
uncertainties.

\section{Results}
\label{res}

\subsection{Full sample}

\begin{figure*}
\begin{center}\parbox{15.3cm}{\hbox{
\psfig{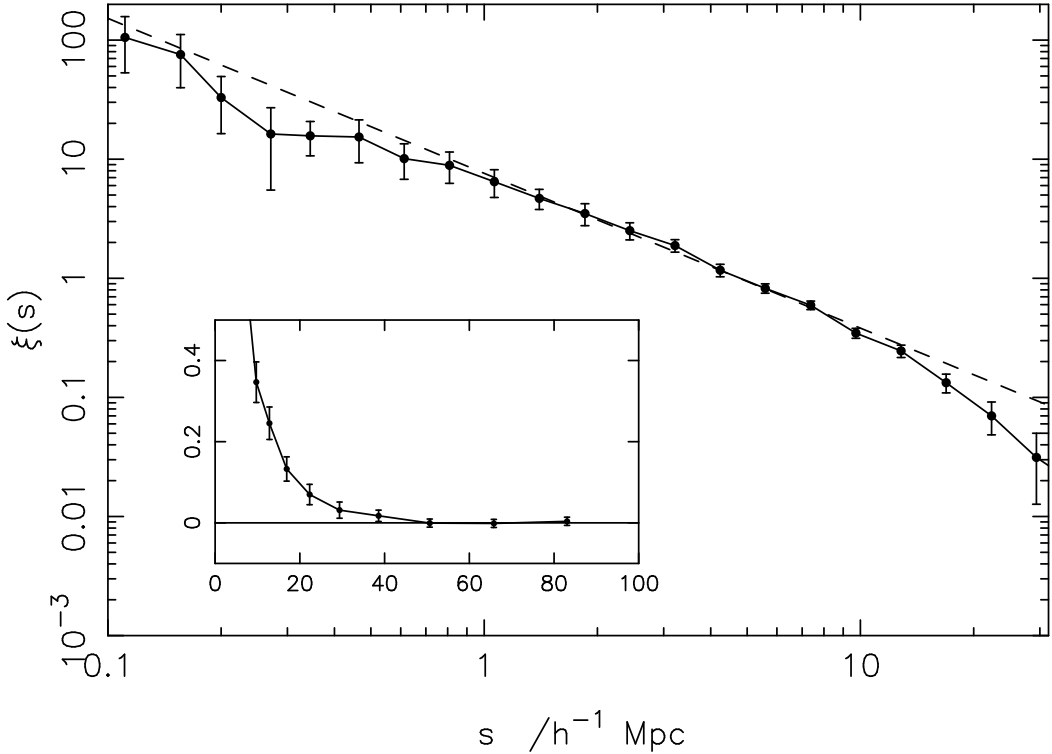}}}
\caption[]{ Logarithmic plot of $\xi$ against $s$ for the full PSCz
sample. The dotted line is the best-fit power-law, fitted for $1 < s <
10 \hmpc$. The inset shows a linear plot of $\xi$ against $s$. The
error bars are derived from the scatter between ten realizations of
the $\Lambda CDM$ mock galaxy catalogues.  }
\label{f_xi_all}
\end{center}
\end{figure*}

Our estimate of $\xi(s)$ for the full sample is shown in
Figure~\ref{f_xi_all}. The power law fit gives $s_0 = 4.77 \pm 0.20$
and $\gamma = 1.30 \pm 0.04$. The errors are estimated from the
standard deviation between the ten realizations of the $\Lambda CDM$ mock
catalogues. We calculated the value of $J_3$ and found it to be $1500
\pm 400$ at a scale of $75 \hmpc$. The error on the $J_3$ result is
from the standard deviation of the mock realization results.

\subsection{Colour sub-samples} 

\begin{figure*}
\begin{center}\parbox{15.3cm}{\hbox{
\psfig{figure=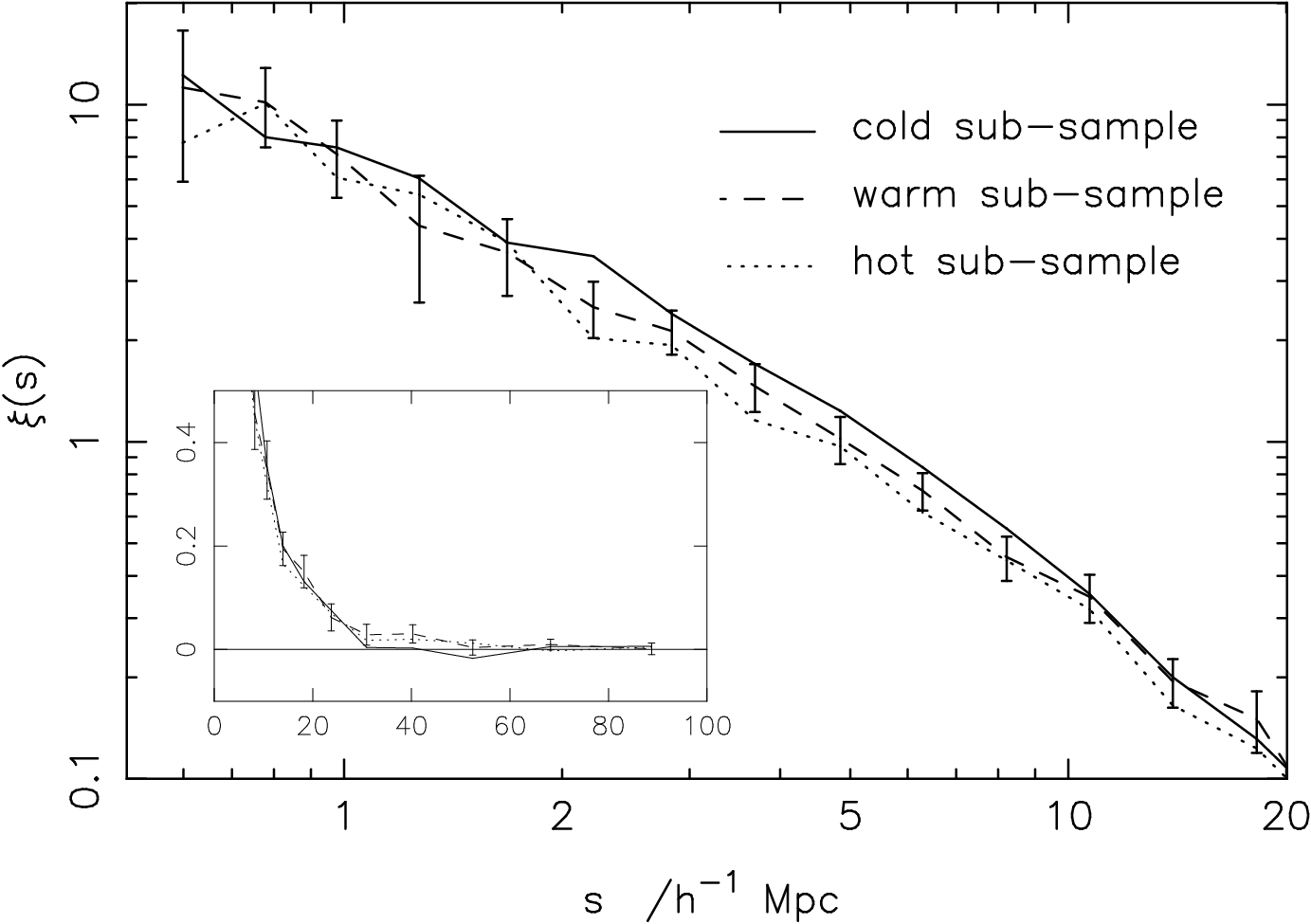,angle=-90,height=10cm, clip=}}}
\caption[]{ Logarithmic plots of $\xi$ against $s$ for the colour
sub-samples defined in Table~\ref{t_fits_col}. The inset shows a
linear plot of $\xi$ against $s$. The error bars are derived from the
scatter between ten realizations of the $\Lambda CDM$ mock galaxy
catalogues. For clarity, error bars are plotted only for the warm
sub-sample.  }
\label{f_xi_col}
\end{center} 
\end{figure*}

\begin{figure}
\psfig{figure=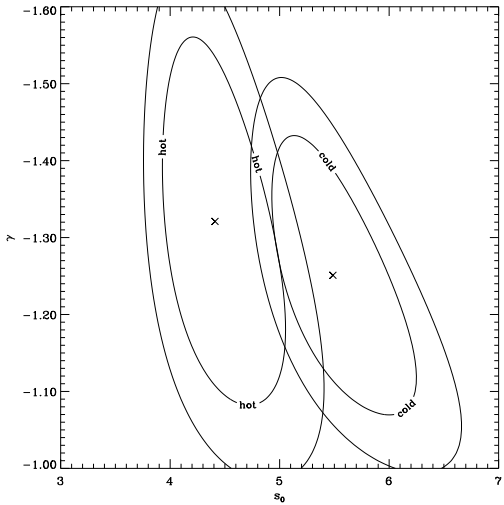,angle=0,height=7cm} 
\caption[]{ Contours of $\chi^2$ as a function of $s_0$ and $\gamma$
for the power-law fits to $\xi$ of the hot and cold sub-samples as in
Table~\ref{t_fits_col}.  
}\label{fig_s0gamma} 
\end{figure}

Our estimates of the redshift-space correlation functions for the
three colour sub-samples are shown in Figure \ref{f_xi_col}.  The
error bar for each point shows the standard deviation between 10
realizations of the $\Lambda CDM$ mock catalogues. It is clear that
over the range $1<s<10~h^{-1}{\rm Mpc}$ the amplitude of $\xi$(s) is
higher by about a factor of 1.5 for the cooler sub-sample.  Fitting a
power law to the data over the range of $1<s<10~h^{-1}{\rm Mpc}$ gives
parameters as shown in Table~\ref{t_fits_col}, clearly confirming the
increase in clustering amplitude for cooler
galaxies. Figure~\ref{fig_s0gamma} shows $\chi^2$ contours for the hot
and cold sub-sample, and suggests that trend is significant at about
the $2\sigma$ level.

Since the best-fit $s_0$ and $\gamma$ are correlated, we have also
fitted the data using a fixed value for $\gamma$, set to the best fit
value for the full sample, $\gamma=1.30$. The resulting $s_0$ values
for the three different colour sub-samples are listed in
Table~\ref{t_fits_col} and plotted as a function of mean colour in
Figure~\ref{f_amp_col}.  The trend of increasing $s_0$ for warmer
galaxies is clearly apparent at about the $3\sigma$ level.  A simple
linear regression to the points yields 
\begin{equation} 
\log_{10}(s_0) = 0.24 \log_{10}( \frac{f_{100}}{f_{60}}) + 0.61.
\label{eqn_s0}
\end{equation}

Analyzing colour selected sub-samples from the mock catalogues shows
no significant difference in amplitude between the hot and cold
sub-samples. This is as expected since we assigned the colours at
random,  but the measurements provide an important check that there
are no subtle biases in our algorithms. 

\begin{table}
\caption[]{ Power law fit parameters to the colour sub-samples } 
\begin{tabular}{@{}lll} 
 Free parameter best fits: \\   
 colour range&  $s_0$ & $\gamma$ \\ 
 $ f_{100}/f_{60} < 1.7$       &   4.41 $\pm$ 0.29  &  1.32 $\pm$ 0.06  \\ 
 $ 1.7 < f_{100}/f_{60}< 2.3$  &   4.76 $\pm$ 0.30  &  1.21 $\pm$ 0.06  \\ 
 $ f_{100}/f_{60} > 2.3$       &   5.49 $\pm$ 0.29  &  1.25 $\pm$ 0.05  \\ 
\\
 Fixed $\gamma$ at 1.30: \\
 colour range&  $s_0$  \\ 
 $ f_{100}/f_{60} < 1.7$       &   4.43 $\pm$ 0.30 \\ 
 $ 1.7 < f_{100}/f_{60}< 2.3$  &   4.64 $\pm$ 0.33 \\ 
 $ f_{100}/f_{60} > 2.3$       &   5.38 $\pm$ 0.30 \\ 
\end{tabular}
\label{t_fits_col} 
\end{table}

\begin{table}
\caption[]{ Power law fit parameters to the luminosity sub-samples } 
\begin{tabular}{@{}lll}
 Free parameter best fits: \\   
 luminosity range&  $s_0$ & $\gamma$ \\ 
 $ log(L_{60}) < 9.6$           &  4.96 $\pm$ 0.44  &  1.21 $\pm$ 0.04  \\ 
 $ log(L_{60}) > 9.6$           &  4.80 $\pm$ 0.31  &  1.36 $\pm$ 0.06  \\ 
\\
 Fixed $\gamma$ at 1.30: \\
 luminosity range&  $s_0$ \\
 $ log(L_{60}) < 9.6$           &  4.82 $\pm$ 0.54 \\ 
 $ log(L_{60}) > 9.6$           &  4.89 $\pm$ 0.35 \\   
\end{tabular}
\label{t_fits_lum} 
\end{table}

\subsection{Luminosity sub-samples}

\begin{figure*}
\begin{center}\parbox{15.3cm}{\hbox{
\psfig{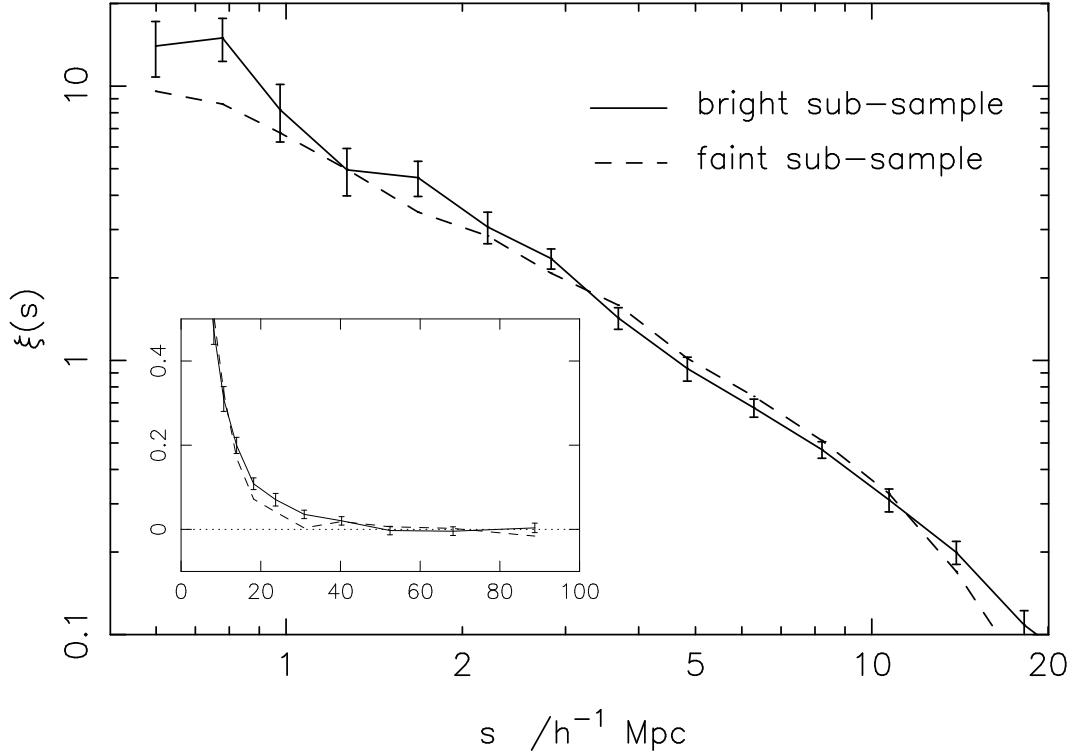}}}
\caption[]{ Logarithmic plots of $\xi$ against $s$ for the luminosity
sub-samples defined in Table~\ref{t_fits_lum}. The inset shows a
linear plot of $\xi$ against $s$. The error bars are derived from the
scatter between ten realizations of the $\Lambda CDM$ mock galaxy
catalogues. For clarity, error bars are plotted only for the bright
sub-sample.  }
\label{f_xi_lum}
\end{center}
\end{figure*}

The estimates for the two luminosity sub-samples are shown in Figure
\ref{f_xi_lum}. The error bars are derived from the scatter between ten
realizations of the $\Lambda CDM$ mock galaxy catalogues. It is clear
that there is very little difference between the two sub-samples. The
power law fit parameters are shown in Table~\ref{t_fits_lum}. These
parameters show no significant difference when the errors are taken
into account.

For all mock catalogues we found no difference between the results for
the faint and bright sub-samples within the errors. This is as
expected since we assigned the luminosities at random, but the
measurements provide an important check that there are no subtle
biases in our algorithms.

\subsection{Volume limited sub-samples}

\begin{figure*}
\begin{center}\parbox{15.3cm}{\hbox{
\psfig{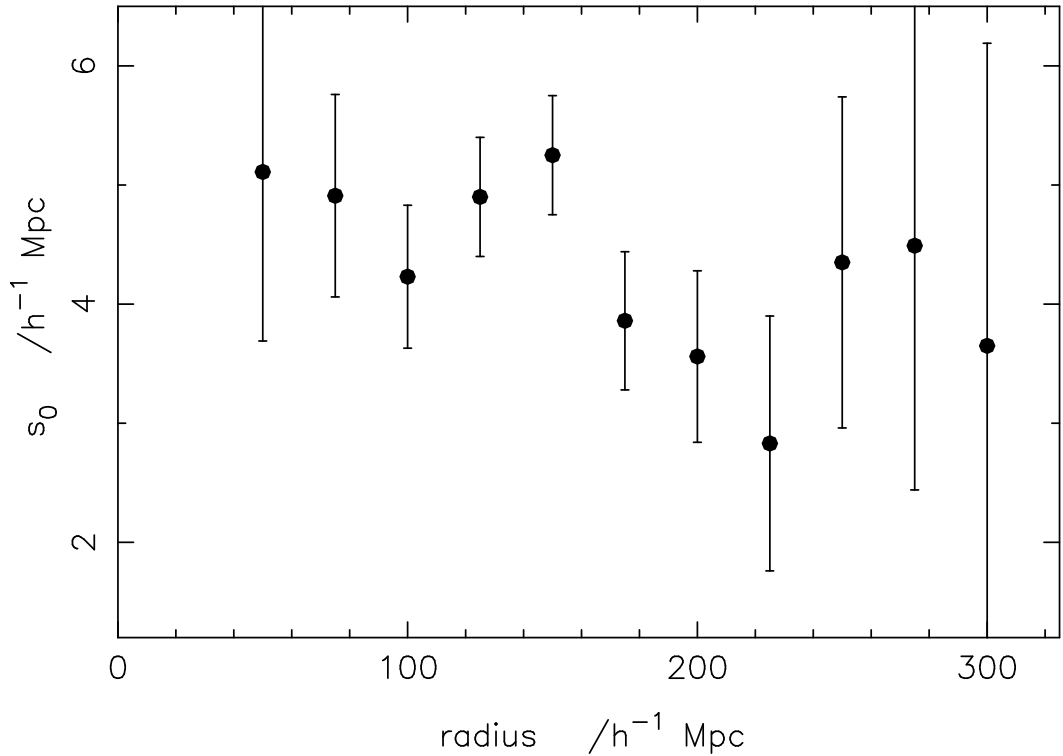}}}
\caption[]{ Clustering length $s_0$ versus volume radius for the
volume limited sub-samples for a fixed $\gamma$ = 1.30. The error
bars show the overall uncertainties including both Poisson errors in the
pair counts, and an estimate of the cosmic variance as described in
Section~\protect\ref{sect_vollim}. For volumes with radius $R<100\hmpc$ we
also estimated the uncertainties from the standard deviation between
the ten realizations of the $\Lambda CDM$ mock catalogues and found
good agreement with the analytic estimates. 
} 
\label{f_vlfit}
\end{center} 
\end{figure*}

Our results from the volume limited sub-samples are shown in
Table~\ref{t:vollim} and displayed in Figure~\ref{f_vlfit}. Although
there is clearly a hint of a trend of reducing amplitude with scale,
it can be seen that the error bars are too large to say that this is a
statistically significant relation. 

The minimum absolute luminosity in the volume limited samples
increases with the radius of the volume, so the mean luminosity also
increases. Figure~\ref{f_vlpred} explicitly shows the clustering
amplitude as a function of the mean luminosity in the volume limited
samples. 

\begin{table}
\caption[]{ Maximum likelihood fits to the volume limited sub-samples
with $\gamma$ fixed at 1.30, and uncertainties estimated as described
in Section~\protect\ref{sect_vollim}. }
\begin{tabular}{@{}cccc}
  Volume radius&  no. of & $\left\langle
\log_{10}\left(\frac{L_{\phantom{\odot}}}{L_\odot}\right)\right\rangle$
&  $s_0$ \\
($h^{-1}{\rm Mpc}$)& galaxies & & ($h^{-1}{\rm Mpc}$) \\  

 50  & 1699 &  9.41 & 5.11 $\pm 1.19 $ \\
 75  & 2151 &  9.71 & 4.91 $\pm 0.65 $ \\
 100 & 2188 &  9.94 & 4.23 $\pm 0.44 $ \\
 125 & 1961 & 10.12 & 4.90 $\pm 0.38 $ \\
 150 & 1633 & 10.28 & 5.25 $\pm 0.42 $ \\
 175 & 1399 & 10.41 & 3.86 $\pm 0.53 $ \\
 200 & 1202 & 10.53 & 3.56 $\pm 0.74 $ \\
 225 & 958  & 10.64 & 2.83 $\pm 1.08 $ \\
 250 & 779  & 10.74 & 4.35 $\pm 1.43 $ \\
 275 & 653  & 10.83 & 4.49 $\pm 2.00 $ \\
 300 & 587  & 10.91 & 3.65 $\pm 2.50 $ \\
\end{tabular}
\label{t:vollim} 
\end{table}

\begin{figure}
\begin{center}\parbox{15.3cm}{\hbox{
\psfig{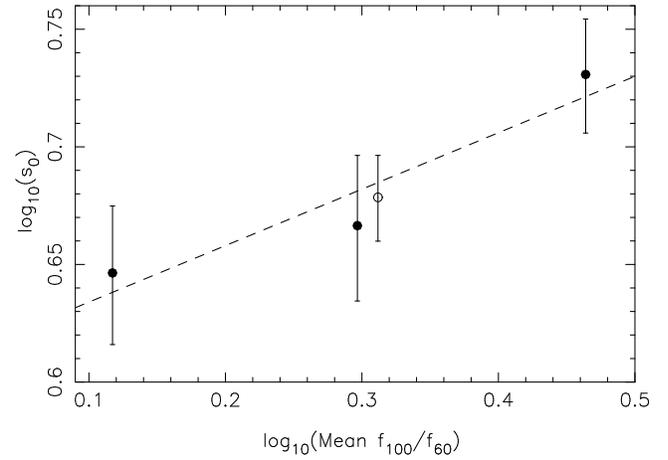}}}
\caption[]{Clustering length $s_0$ versus mean colour for $\gamma$ =
1.30. The filled points show the colour sub-samples as defined in
Table~\ref{t_fits_col}, and the open point shows the full catalogue. The best fit line is also shown and has the form $\log_{10}(s_0) = 0.24 \log_{10}(
\frac{f_{100}}{f_{60}}) + 0.61$.} 
\label{f_amp_col}
\end{center} 
\end{figure}

\section {Conclusions and discussion}
\label{discuss}

We detect a colour dependence in the clustering of {PSCz galaxies: the
cooler galaxies are more strongly clustered than the hotter galaxies.
Although the significance level of amplitude variations is not high,
this does not mean that the effects are small: the clustering
amplitude differs by a factor of 1.5 between the warm and cool
sub-samples.  We do not detect any significant variation in clustering
for luminosity selected subsamples. Volume limited sub-samples show a
slight trend corresponding to a weaker clustering for brighter
galaxies, but the uncertainties mean it is not significant.

Our results are different to those of Mann, Saunders \& Taylor (1996)
who found marginal evidence that warmer galaxies are more strongly
clustered than cooler galaxies. The main reason for the difference is
simply the factor of 6 increase in number of galaxies in our sample,
which dramatically reduces the uncertainties in our clustering
estimates.  The uncertainties on the earlier measurements encompass
our results, so there is actually no discrepancy.

The variation of clustering with FIR colour that we find in the PSCz
is consistent with the variation with star-formation rate indicated by
H$\alpha$ and \oii emission lines in the SAPM survey (Loveday \etal
1999).  Loveday \etal split the SAPM survey into three samples with
low, medium and high H$\alpha$ equivalent widths (EW), and estimate
$s_0$ for the samples to be 8.7, 5.5 and 4.6 $h^{-1}{\rm Mpc}$
respectively.  A similar split into low, medium and high \oii EW
samples gives $s_0$ values of 8.6, 4.9 and 4.1 $h^{-1}{\rm Mpc}$
respectively.  The low EW samples are early type galaxies, which do
not appear in IRAS samples. The medium and high EW samples are
actively star-forming galaxies, which are more similar to IRAS
samples. The clustering amplitude of the medium and high EW galaxies
is very similar to that of the IRAS galaxies, suggesting that they do
indeed trace a similar population.  Furthermore, the change in $s_0$
between the medium and high SAPM emission-line galaxies is very
similar to the change between cold and hot PSCz galaxies as seen in
Figure~\ref{f_amp_col}.

The luminosity dependence is a more complex effect, because the FIR
luminosity of a galaxy depends on both the mass of dust, and the
recent star-formation rate. If the dust mass is correlated with the
halo mass, then more luminous galaxies will, on average, be in more
massive halos, which are expected to have a higher clustering
amplitude. On the other hand, a high star-formation rate will make a
galaxy more luminous, and galaxies with a high star-formation rate
tend to be less clustered. It is not obvious which will be the
dominant effect.

Given the correlation between colour and luminosity seen in
Figure~\ref{f_colf} we can infer the luminosity dependence from the
colour dependence seen in Figure~\ref{f_amp_col}. From
Figure~\ref{f_colf} we find $\log_{10}(\frac{f_{100}}{f_{60}}) =
-0.117\log_{10}(L_{60}) + 1.427$, and so together with
equation~\ref{eqn_s0} we would expect $\log_{10}(s_0) = -0.028
\log_{10}(L_{60}) + 0.95$. This relation is plotted as the dashed line
on Figure~\ref{f_vlpred}, and it can be seen that it is consistent
with the data points.  The luminosity variation seen in optical galaxy
samples is approximately of the form $\log_{10}(s_0) =
4.37(0.7+0.3\frac{L}{L^*})$ (Benoist \etal 1996). The dotted line
shows this with $L^* = 3.6\times 10^9 L_\odot$ (Springel \& White 1998).
This is clearly inconsistent with the trend seen in the PSCz sample.

The small decrease in clustering that we find for more luminous
galaxies is consistent with the colour dependence effect alone, with
no evidence for an increase in clustering for more luminous galaxies.
If the standard picture of high-peak biasing is correct, we conclude
that there is a very weak correlation between the FIR luminosity of a
galaxy and the mass of it's dark-matter halo. 
However, models which allow a variable number of galaxies per dark halo,
and include scatter about the mean mass-luminosity relation lead to a
very weak luminosity dependence of clustering (Somerville \etal 1999).

Finally we note that our present analysis has been restricted to
the redshift space correlation functions. This means that any
differences in the velocity distributions of the different sub-samples
of galaxies will affect the amplitude of $\xi(s)$, particularly on
small scales where the velocity separations are comparable to the
random peculiar velocities. 
We consider these effects in a future paper. 

\begin{figure*}
\begin{center}\parbox{15.3cm}{\hbox{
\psfig{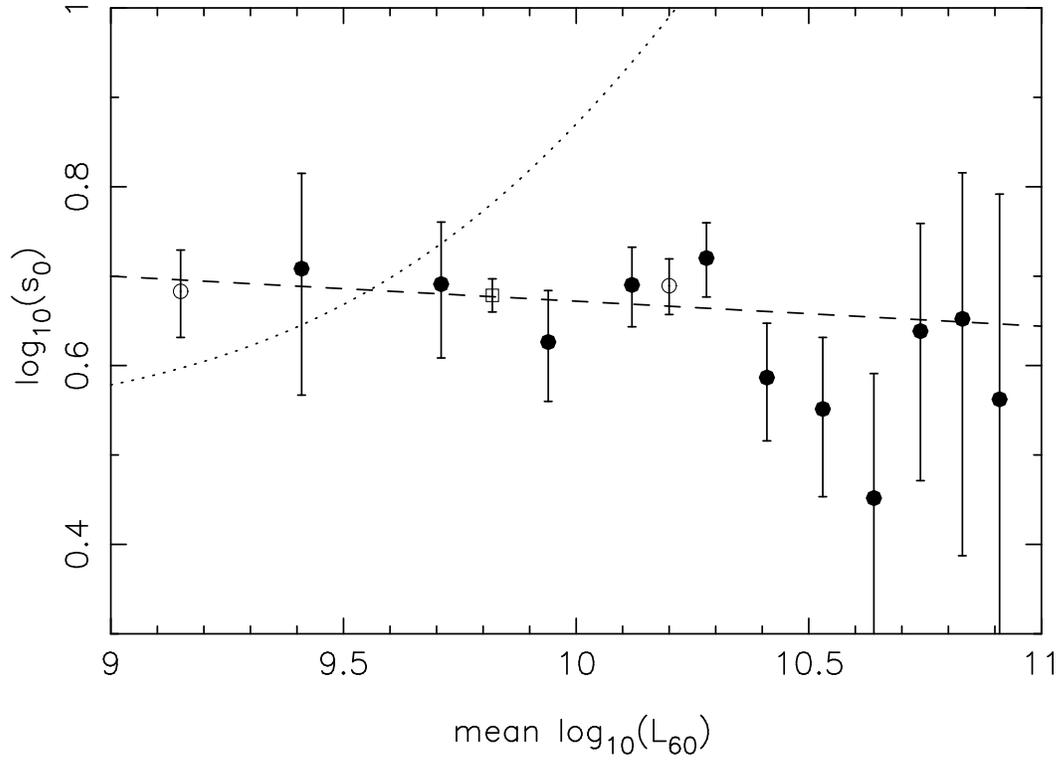}}}
\caption[]{ Log plot of clustering length $s_0$ versus mean absolute
luminosity for the volume limited sub-samples (filled circles), the
bright and faint sub-samples (open circles) and the whole sample (open
square). The dashed line shows the trend expected from the colour
dependence seen in Figure~\ref{f_amp_col} and the luminosity-colour
correlation in Figure~\ref{f_colf}. The dotted line shows
$\log_{10}(s_0) = 4.37(0.7+0.3\frac{L}{L^*})$, which approximates the variation
seen in optical galaxy samples.}
\label{f_vlpred}\end{center} 
\end{figure*}

\section*{Acknowledgements} 

We would like thank everyone who has contributed to the construction
of the PSCz survey, and also the referee (Andrew Hamilton) for
suggesting many useful improvements to the original draft.

\section*{References} 

\refer Beichman C., Neugebauer G., Habing H.J., Clegg P.E.,
Chester T.J., 1988, IRAS Catalogs and Atlases Explanatory Supplement, 
NASA RP-1190 vol 1, US Government Printing Office, Washington DC. 

\refer Beisbart Claus, Kerscher Martin, 2000, \ApJ~accepted, astro-ph/0003358.

\refer Benoist, C., Maurogordato, S., da Costa, L.N., Cappi, A.,
Schaeffer, R., 1996, \ApJ, 472, 452. 

\refer Cole S., Hatton S., Weinberg D., Frenk C., 1998, \MN, 300, 945.

\refer Couchman H.M.P., 1991, \ApJ, 368, L23.

\refer Croft R., Dalton G., Efstathiou G., Sutherland W., Maddox S.,
1997, \MN, 291, 305.  

\refer Dekel A., Rees M.J., 1987, \Nature, 326, 455.

\refer Dressler A., 1980, \ApJ, 236, 351.

\refer Efstathiou G., 1988, in Proc. 3rd IRAS Conf., Comets to
Cosmology, ed. A.Lawrence (New York:Springer), 312. 

\refer Hamilton A.J., 1997a, \MN, 289, 285.

\refer Hamilton A.J., 1997b, \MN, 289, 295.

\refer Kaiser, N., 1984, \ApJ, 284, L1. 

\refer Katz N., Gunn J.E., 1991, \ApJ, 377, 365.

\refer Landy S.D., Szalay A.S., 1993, \ApJ, 412, 64.

\refer Lawrence, A., Rowan-Robinson, M., Ellis, R. S., Frenk, C. S.,
Efstathiou, G., Kaiser, N., Saunders, W., Parry, I. R., Xiaoyang, Xia,
Crawford, J., 1999, \MN, 380, 897. 

\refer Loveday J.,  Maddox, S.J., Efstathiou G., Peterson B.A., 1995,
\ApJ, 442, 457.

\refer Loveday J., Tresse L., Maddox S., 1999, \MN, 310, L281.

\refer Mann R.G., Saunders W., Taylor A.N., 1996, \MN, 279, 636

\refer Mo, H.J., White, S.D.M., 1996, \MN, 282 347. 

\refer Moore B., Frenk C.S., Efstathiou, G.P., Saunders, W., 1994,
\MN, 269, 742. 

\refer Navarro, J.F., White, S.D.M., 1993, \MN, 265, 271.

\refer Park C., Vogeley M.S., Geller M.J., Huchra J.P., 1994, \ApJ, 431, 569. 

\refer Pearce F.R. et al., 1999, \ApJ, 521, 99.

\refer Rosenberg Jessica L., Salzer John J., Moody J. Ward, 1994, \AJ,
108, 1557. 

\refer Saunders, W., Sutherland, W. J., Maddox, S. J., Keeble, O.,
 Oliver, S. J., Rowan-Robinson, M., McMahon, R. G., Efstathiou, G. P.,
 Tadros, H., White, S. D. M., Frenk, C. S., Carrami\~nana, A.,
 Hawkins, M. R. S., 2000, \MN 317, 55.

\refer Saunders, W., Rowan-Robinson, M., Lawrence, A., Efstathiou, G.,
 Kaiser, N., Ellis, R. S., Frenk, C. S., 1990 \MN 242, 318.

\refer Somerville, R.S.,  Lemson, G.,  Kolatt, T.S.,  Dekel, A., 
1999, astro-ph/9912073

\refer Soifer, B. T., Sanders, D. B., Madore, B. F., Neugebauer, G.,
Danielson, G. E., Elias, J. H., Lonsdale, Carol J., Rice, W. L., 1987,
\ApJ  320, 238. 

\refer Springel, V., White, S.D.M., 1998, \MN, 298, 143. 

\refer Szapudi Istv${\rm \acute{a}}$n, Branchini Enzo, Frenk C.S.,
Maddox Steve, Saunders Will, 2000,  MNRAS, 2000, 318 L45. 

\refer Valls-Gabaud D., Alimi J.-M., Blanchard A., 1989, \Nature, 341, 
215.

\refer Vogeley M.S., Szalay A.S., 1996, \ApJ, 465, 34. 

\refer White S.D.M., Rees M.J., 1978, \MN, 183, 341.

\end{document}